# Microscale deformation of intermetallic-Mg interface under shear loading

Anwesha Kanjilal[1,*], Shamsa Aliramaji[2], Deborah Neuß[2], Marcus Hans[2], Jochen M. Schneider[2], James P. Best[1], and Gerhard Dehm[1,*]

[1] Max Planck Institute for Sustainable Materials, 40237 Düsseldorf, Germany

[2] Materials Chemistry, RWTH Aachen University, 52074 Aachen, Germany

[*] Corresponding authors: a.kanjilal@mpie.de, g.dehm@mpie.de

**Abstract**

While intermetallic (IM)-metal interfaces in metallic alloys are critical for tuning mechanical properties, they can also act as failure sites, underscoring the importance of determining their strength. This study reports on a novel microshear geometry, and demonstrates its applicability for testing the strength and deformation behavior of IM-metal interfaces in Mg-Al-Ca alloys, a key material for light weight automotive applications. The shear tests are applied to a model bi-layered system grown by magnetron sputtering, comprising of a $CaMg_2$ film deposited onto a Mg layer. A parametric study was performed using finite element modeling to optimize the specimen dimensions. Subsequently, *in situ* microshear tests conducted inside a scanning electron microscope revealed an interface shear strength of ~136 $\pm$ 12 MPa, and provided insights into the stages of deformation progression. *Post mortem* examination of the sheared interface revealed an irregular surface indicating ductile deformation at room temperature.

**Keywords:** Intermetallic-Mg interface; Interface strength; Microshear testing; *In-situ* mechanics

Interfaces in materials dictate important mechanical properties such as strength, toughness or creep resistance. Continuous efforts have been made to tailor interfaces in advanced materials to further enhance these properties ranging from intermetallic (IM)-metal hetero-interfaces in precipitate-strengthened alloys [1], to higher grain boundary (GB) areas in nanocrystalline metals [2,3], fiber-matrix interfaces in composites [4], or nanoscale distribution of hetero-interfaces in multilayered or lamellar microstructures [5,6]. The most striking feature of interfaces is their ability to influence microscopic deformation processes through local plasticity mechanisms (*e.g.* dislocation pile-up or transmission at GBs) [5,7] and events like crack arrest, deflection, or propagation [8,9]. However, interface failure due to debonding, cracking, or delamination because of insufficient interfacial strength and toughness in application environments, poses a significant challenge which restricts the long-term reliability of advanced materials [5,6,10–12].

In the context of metallic alloys, their behavior is closely associated with the mechanical response of IMs and IM-metal interfaces. One important class of metallic alloys which have gained attention for lightweight automotive applications are the Mg-Al-Ca alloys. They contain an interconnected network of various strengthening IMs such as $\beta$-$Mg_{17}Al_{12}$, $CaMg_2$, $CaAl_2$ or $Ca(Mg,Al)_2$ [10,13]. Damage at these IM-Mg interfaces can arise from stress concentration due to pile-up of lattice dislocations at the hetero-interfaces resulting in fracture of the IM [14,15], IM-metal interface decohesion-cracking [14,15], or plasticity mediated interface shear [10,15,16]. Whether the IM fractures or the interface undergoes plastic deformation by shearing, will depend on the relative strength of each failure site. Therefore, it becomes imperative to determine the strength and toughness of these failure sites *a priori*. Despite longstanding research on Mg alloys, the strength and underlying deformation mechanisms of IM-Mg interfaces are not well established, although these sites undergo local damage under load due to slip transmission, local decohesion and cracking at intermetallic-Mg interface as observed during tensile deformation [15]. The fracture toughness of some of these brittle IMs have, however, been investigated using micromechanical testing [17,18]. Recent atomistic simulations provide additional insights into the local deformation mechanics of $CaMg_2$-Mg interfaces [16] and other IM-metal systems [19,20] at the atomic-level using interface models. In a different study, nanoindentation was used to compare the strain rate sensitivity of the Mg matrix with that of the $Ca(Mg,Al)_2$-Mg interface in an indirect manner [21]. However, a *direct quantitative measurement* of the local interfacial strength of the interfaces is not yet reported. This can be attributed to experimental challenges due to the curved morphology of most IM-metal interface boundaries and nano- to microscale dimensions of the IMs which makes traditional interface strength testing methods, such as fiber pull-out testing [22], indentation testing, scratch testing, bulge testing, 4-point bend testing, or testing of sandwiched layers [8,23], unsuitable for a direct site-specific microscale measurement of the interface strength. The aforementioned bulk-scale methods are also generally not suitable for brittle systems which break down where significant plasticity is present (*i.e.*, nanoindentation or scratching), while complex elastic-plastic stress fields and mixed-mode loading or unwanted buckling often make direct extraction and interpretation of interface strength challenging, resulting in low success rates [8,24].

Recently, a double microshear geometry has been employed locally at the microscale to determine the mode II interface strength and fracture toughness between SiC fiber-pyrolytic interface for SiC carbide reinforced SiC matrix with pyrolytic interphase [12]. The advantage of this technique over those used in the past to investigate interface strength or toughness is the direct and simple quantification of the shear strength also evident from microshear testing on bulk single crystals or composites [12,25]. Hence, this method is largely unaffected by the above-mentioned challenges which exist for macroscale testing methods. However, a continuous boundary is required spanning several micrometers in 3-dimensions between the two phases and symmetric orientation of the interfaces on either side of the applied load in order to conduct a reliable test for interfacial strength and toughness. These conditions are difficult to achieve in Mg alloys where IMs are generally sub-micrometer size with elongated shape and are randomly dispersed with irregular boundaries with the Mg matrix [15], and cannot be used for real alloy systems. Accordingly, in the present study a novel approach is developed which uses shear testing to determine the IM-Mg interface strength at the microscale. A model interface is employed, using a layered $CaMg_2$-Mg film with a continuous and straight interface to mimic IM-Mg interfaces in bulk Mg alloys. $CaMg_2$ is a *C*14 Laves phase intermetallic important for enhancing the strength and creep resistance of Mg alloys [13]. Further, a single shear micro-geometry is employed suitable for thin film configurations, with dimensions optimized by finite element modeling (FEM) to obtain geometric size-independent interface strength values.

A rectangular microscale single shear specimen geometry, as schematically shown in **Fig. 1a**, was used to measure the $CaMg_2$-Mg interface strength. FEM was first performed using quasi-static room temperature material parameters of $CaMg_2$ and Mg. The goal of FEM was to obtain geometric dimensions which have minimum influence of out-of-plane bending, and accurately map the applied load to the load at the interface for interface strength values. In the model, the CaMg$_2$ intermetallic was treated as elastic material [26], while Mg was considered as elastic-plastic material [27]. Due to symmetry along the center, only half of the geometry was modeled (**Fig. 1b**), with symmetric boundary conditions (BCs) applied at the center and all degrees of freedom constrained at the backside of Si to represent a fixed end. Loading was simulated by applying a fixed displacement, $\delta$, on the top surface of CaMg$_2$, parallel to the interface, mimicking experimental conditions wherein the load is applied with indenter on the surface of intermetallic (see **Fig. 3b**). Hexahedral elements were used to discretize the geometry, with mesh refinement at the interface to precisely resolve the deformation across it. **Fig. 1b** shows the 3D discretized geometry with BCs. The geometric parameters (*i.e.*, the length *L*, width *W*, thickness *H*) of the shear ligament are annotated in the figure. **Fig. 1c** depicts the deformed configuration near the interface post-shearing. As the objective of FEM was to optimize specimen dimensions for subsequent interface strength testing, interface debonding or cracking was not considered.

FE analysis revealed that the load parallel to the interface is equal to the applied load, $P_{app}$, on the CaMg$_2$ surface (see **Fig. S1** of the Supplementary Information). Hence, the interface shear stress was calculated as $\tau_{int} = P_{app}/A$, where *A* is the cross-sectional area of the shear ligament. The parametric study varying shear ligament dimensions (*L*, *W*, *H*), as shown in the contour plots of **Fig. 1d-e**, indicated that the interface strength is less sensitive to changes in shear ligament dimensions for thicker specimens (*H*≈5.5 µm or higher) and shorter ligament length (*L*~4 µm or lower). These dimensions ensure uniform shear displacement at the interface with minimum bending effects, as shown in the deformed specimen configuration in **Fig. 1c**.

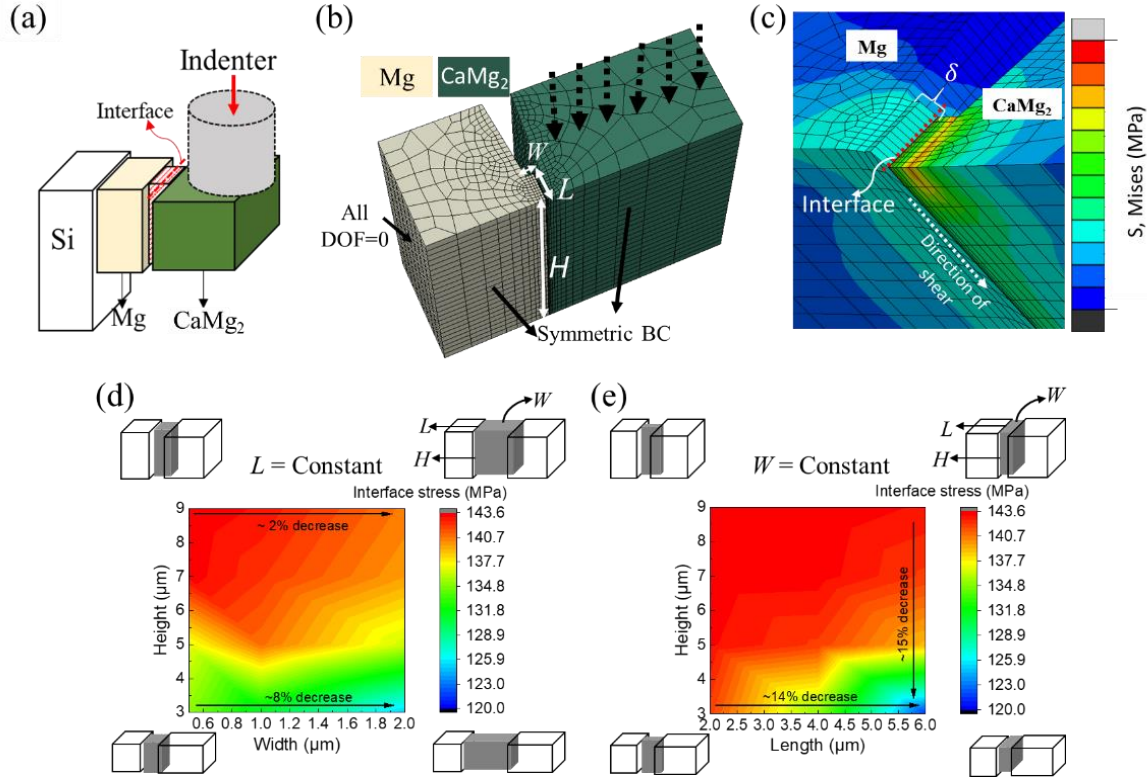

**Fig. 1**: Geometry optimization by finite element modeling in ABAQUS. (a) Schematic representation of the single-shear specimen geometry. (b) Discretized geometry used for FE analysis showing the boundary conditions; half the geometry was modelled due to existing symmetry, (c) Shear deformed specimen. (d-e) Parametric study of shear ligament dimensions on interface strength obtained from FE analysis; contour plots show the effect of variation in (d) width and height and (e) length and height of shear ligament; for visualization the length $L$, width $W$ and height $H$ of the shear ligament are annotated in (b,d,e).

The bilayered CaMg$_2$-Mg model films for interface shear testing were prepared using combinatorial magnetron sputtering with elemental targets of Ca and Mg [28]. A Ca-Mg film was deposited on a Mg layer, grown on a (001) oriented Si wafer at 100 °C, in a high vacuum chamber pumped to base pressure of $2 \times 10^{-5}$ Pa and Ar pressure of 0.5 Pa. The resulting compositionally graded Ca-Mg layer included the composition of interest *i.e.*, CaMg$_2$ intermetallic on Mg layer. Further details of the sample fabrication technique and identification of the region with CaMg$_2$ intermetallic are given in the Supplementary Information. Microstructural and chemical composition analysis was conducted on the cross-section of bilayered thin film by extracting a cross-section region prepared by plasma focused ion beam (PFIB) milling using Xe ions in a Helios 5 Hydra UX dual-beam microscope (Thermo Fisher Scientific, Waltham, MA, USA) equipped with energy dispersive spectroscopy (EDS) detector. EDS line scans were done with an Octane Elite Plus system (EDAX Inc., Mahwah, NJ, USA) at an acceleration voltage of 8 kV. The microstructure and composition of the film cross-section are shown in **Fig. 2**. The compositions matched expected values and there was no significant oxygen incorporation at the interface.

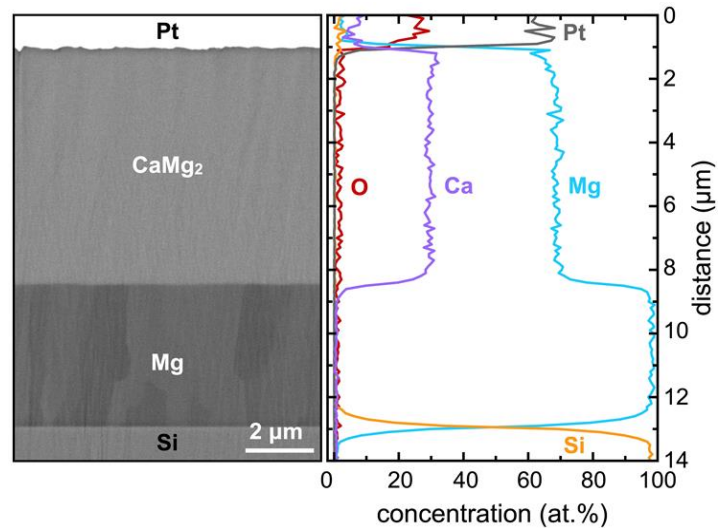

**Fig. 2**: Back scattered electron image showing the microstructure of the cross-section of CaMg$_2$-Mg film on Si substrate used in this study along with the corresponding composition analysis by EDS.

Microshear tests were conducted at room temperature to determine the CaMg$_2$-Mg interface strength and deformation behavior. Specimens were fabricated using Ga-FIB (Zeiss Auriga dual-beam scanning electron microscope (SEM) following the schematic shown in **Fig. 1a**. A trench was first prepared to obtain a sufficiently large region having straight flat edge of the CaMg$_2$-Mg-Si interface suitable for the shear specimen fabrication. Two step milling was performed using 2 nA ion beam current for coarse milling of the rectangular shear specimens, followed by 600 pA beam current for fine milling of the straight edges of the specimens. Subsequently, two grooves (~3.5 µm apart) were milled at the interface to localize the deformation within the shear ligament. Seven specimens were prepared, with three having a height $H \approx 5.8$ µm and four with $H \approx 7.8$ µm to assess the effect of specimen dimensions on the interface strength. All specimens had shear ligament length of ~3.5 µm and width ~0.5 µm, within the FE-optimized range of specimen dimensions, to ensure minimum influence of out-of-plane bending and changes in shear ligament dimensions on interface deformation. *In situ* shear tests were conducted in Hysitron PI-88 (Bruker Inc.) nanoindentation system inside Zeiss Gemini SEM. A 10 µm diameter tungsten carbide flat punch indenter (Synton-MDP, Switzerland) was used to apply load uniformly on the top surface of the CaMg$_2$ intermetallic. The tests were conducted in quasistatic pseudo-displacement-controlled mode at a displacement rate of 10 nm/s. Post-testing the deformed specimens were examined using SEM to analyze the sheared interface and surrounding deformation features.

**Fig. 3a** shows representative microshear test specimens fabricated by FIB, while **Fig. 3b** shows *in situ* loading of a specimen prior to testing inside the SEM. The CaMg$_2$-Mg interfaces are marked with yellow arrows. **Fig. 3c,d** show the representative load-displacement response during shear testing of specimens with thickness $H \approx 5.8$ µm and 7.8 µm, respectively. No noticeable difference in the overall behavior is observed between the two, apart from higher loads needed to deform the thicker specimens. After an initial linear behavior, the load continues to increase with displacement until it reaches a peak value, before decreasing steadily with further deformation. The corresponding change in interface shear stress (calculated as $\tau_{int} = P_{app}/A$ , according to the procedure used in FE analysis) with displacement are also shown in **Fig. 3c,d**.

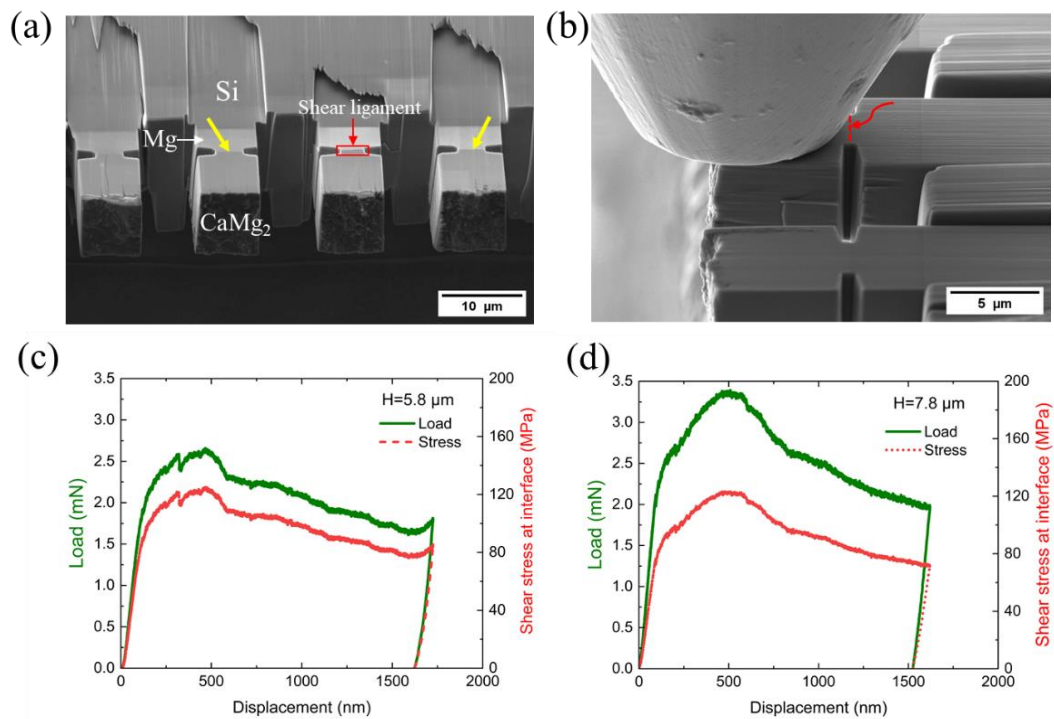

**Fig. 3**: CaMg$_2$-Mg interface shear test: (a) FIB machined micro-shear specimens, and (b) *in situ* loading inside the SEM prior to testing; variation of load and engineering shear stress at the interface with displacement for (c) 5.8 µm and (d) 7.8 µm thickness. The yellow arrows in (a) and (b) indicate the CaMg$_2$-Mg interface. The small load increase before unloading in (c) is due to slight contact of the side of indenter with top edge of Mg.

**Fig. 4a** shows a representative deformed specimen after testing, with shearing occurring at the CaMg$_2$-Mg interface without bending effects. The shearing process is accompanied by cracking at the interface (shown by solid yellow arrow) and plasticity in Mg (region indicated by dashed white arrow). The sheared surface shown in **Fig. 4b** appears rough and irregular, indicating that the interface deformed in a ductile manner with plasticity. It is to be noted that before interface shearing began, plastic flow caused some curving of the Mg material at the top part of interface, but no visible out-out-plane deformation of the interface occurred consistent with FE analysis. SEM-EDS composition analysis performed on the top surface across the sheared interface in FEI-Scios (Thermo Fisher Scientific) SEM further indicated that shearing occurred near the $CaMg_2$-Mg interface (refer to **Fig. S3** of the Supplementary Information).

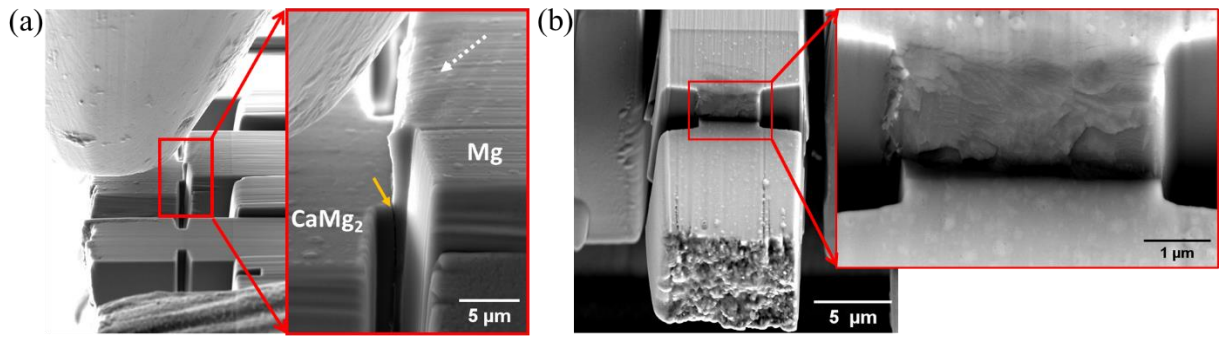

**Fig. 4**: Micrographs showing (a) side view of the shear deformed $CaMg_2$-Mg interface with the interfacial crack denoted by solid yellow arrow, (b) front view of the sheared surface with inset showing the corresponding high magnification image of the sheared interface zone.

Correlating the load-displacement behavior with progressive shearing observed from the *in situ* SEM images reveals different stages of deformation at the interface as shown in **Fig. 5a.** Three different regimes can be identified from the time-dependent deformation analysis: (i) *Perfect bonding* (**Fig. 5a-c**) where the interface remains intact, exhibiting linear load-displacement behavior indicative of overall elastic deformation; (ii) *Sticking* (*i.e.*, a transition between bonding and sliding **Fig. 5d-e**) where the load-displacement curve deviates from linear behavior due to plasticity in Mg adjacent to the interface. A surface step forms at the interface near peak load (marked by arrow in **Fig. 5d,e**) followed by a load drop and indicates the onset of interface shearing; (iii) *Interface sliding* (**Fig. 5f-i**) occurs as the load progressively decreases with deformation at the interface due to geometric effect. Thus, the $CaMg_2$-Mg interface starts shearing near the peak load corresponding to a maximum interface strength of $\sim 124 \pm 10$ MPa for specimens with $H \approx 5.8$ µm and $136 \pm 12$ MPa for specimens with $H \approx 7.8$ µm, with a difference of ~8% within the scatter of experimental data. The transition from bonding to sliding is facilitated by plastic deformation of Mg near the interface, while the $CaMg_2$ IM does not show any visible deformation features. Interestingly, contrary to the brittle behavior of individual IMs in Mg-Al-Ca alloys [17,18], the CaMg$_2$-Mg interface deforms in a ductile manner.

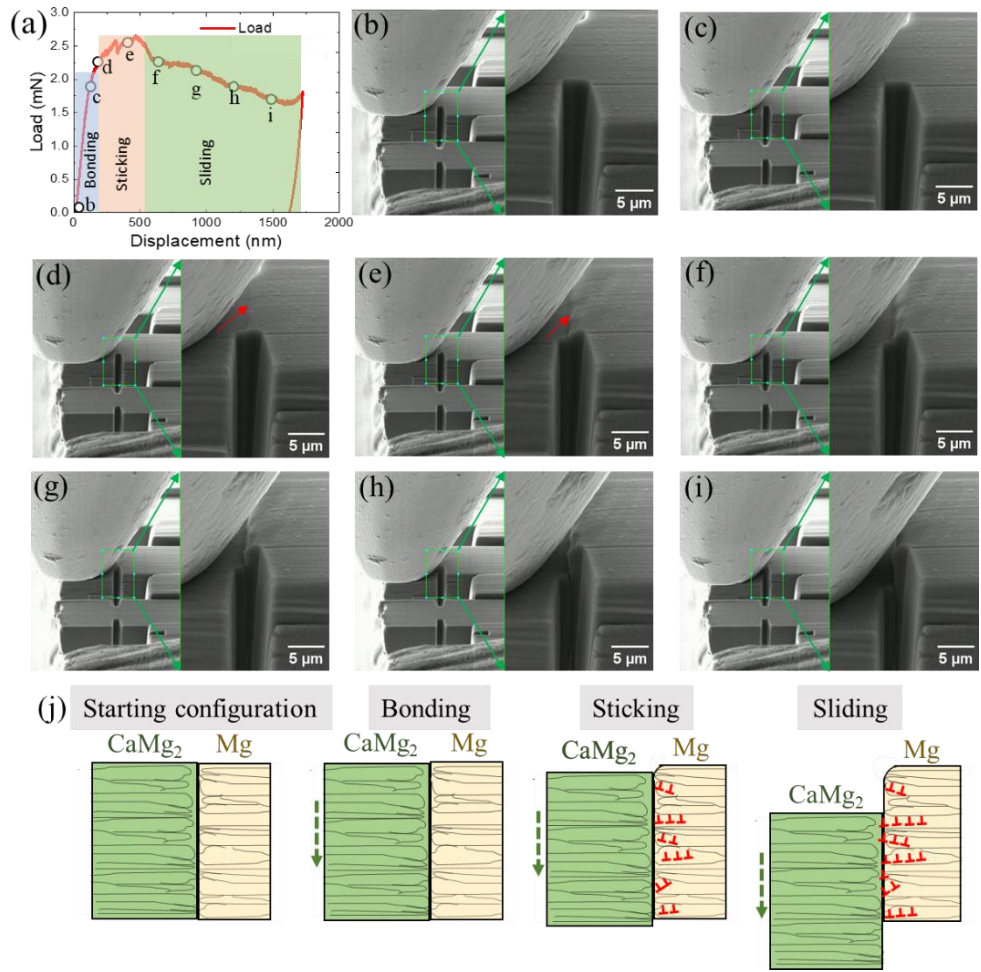

**Fig. 5**: Time-dependent interface deformation analysis correlating (a) load-displacement data with (b-i) different stages of interface deformation from *in situ* testing. (j) Schematic illustration of the stages of interface shearing and deformation mechanism due to dislocation pile-up and sliding at the CaMg$_2$-Mg interface.

Plasticity in Mg can lead to dislocation pile-ups creating stress concentrations at the CaMg$_2$-Mg interface. These stresses may be released through slip transmission across the interface depending on the orientation relationship between CaMg$_2$ and Mg, or by dislocation absorption at the interface that promotes interface sliding [16] (see schematic representation, **Fig. 5j**). Similar mechanisms of Mg plasticity and dislocation pile-up mediated IM-Mg interface decohesion and sliding have been reported in bulk Mg-Ca-Al alloys [15], and from molecular dynamics (MD) simulations of CaMg$_2$-Mg interfaces [16]. The simulations showed that CaMg$_2$-Mg interface sliding occurred when the applied load favored sliding in planes parallel to interface for certain geometries, facilitated by dislocation absorption at the interface from pile-ups in Mg [16]. This behavior was independent of interface orientation relationship and strength of the pile-up. This suggests that for the $CaMg_2$-Mg interface, the insights on interface sliding behavior obtained from fine-grained thin films (where different dislocation pile-ups would be activated in the fine Mg grains at the interface of a given shear specimen, refer to **Fig. 5j**) could potentially be compared to bulk alloys (where a given $CaMg_2$-Mg interface has a distinct orientation). Nonetheless, such extrapolations must be made with caution, and cannot be generalized for other material systems.

The determined strength of elastic-plastic $CaMg_2$-Mg interface is comparable to $SiO_2$-Cu interface for which maximum strength of ~200 MPa was reported with elastic deformation in both materials [24]. For an intermetallic-intermetallic boundary such as FeAl-$FeAl_2$ interface, the interfacial shear strength was ~350 MPa and facilitated easy dislocation glide [29]. For Fe-precipitate interfaces, MD simulations revealed an interface strength less than 10 MPa with dislocation interactions active across the interface [19]. Furthermore, a comparison with various fiber composites shows that their interface strengths typically range from 20 to 70 MPa, depending on fiber dimensions and composite chemistry [4,30]. Additionally, the $CaMg_2$-Mg interface strength is comparable to the tensile yield strength of ultra-fine-grained Mg thin films (~130 MPa) [27]. Since no pre-crack was used in this study, fracture toughness or fracture energy was not determined. However, the mode II energy area density for the $CaMg_2$-Mg interface was estimated using the method employed for a Cu-$SiO_2$ interface [24]. By normalizing the area under the force-displacement curve up to the onset of shearing by the sheared area, an average energy area density of $44 \pm 2$ $J/m^2$ was calculated for the $CaMg_2$-Mg interface. To put this value in perspective, the mode II interface energy density of Cu-$SiO_2$ interface is ~20 $J/m^2$ [24], while metal-ceramic interfaces (*e.g.*, Cu, Al, Au on various substrates) have fracture energies ranging from 0.3 to about 100 $J/m^2$ [31,32], and fiber-matrix interfaces generally fall within 1-50 $J/m^2$ [4,30].

In conclusion, using a novel single microshear testing approach on a model thin film system the strength of $CaMg_2$-Mg interface was determined for the first time, with a maximum shear strength of ~136 MPa achieved using FEM-optimized specimen dimensions. The tests revealed different stages from perfect bonding to sticking and finally sliding of the interface, driven by plasticity in Mg near the interface. This geometry will enable systematic *in situ* micromechanical studies on interface shearing, including as a function of temperature or strain rate as well as for hetero-interfaces in thin film systems.

**Acknowledgement**

The authors acknowledge the financial support from the Deutsche Forschungsgemeinschaft (DFG) for projects B06, B03 and B02 within Collaborative Research Center (SFB) 1394 "Structural and Chemical Atomic Complexity - from defect phase diagrams to material properties", project number 409476157.

**Conflict of interests or competing interests**

The authors declare that no conflict of interest or competing interests exist.

*Supplementary Information for:*

**Microscale deformation of intermetallic-Mg interface under shear loading**

Anwesha Kanjilal[1],*, Shamsa Aliramaji[2], Deborah Neuß[2], Marcus Hans[2], Jochen M. Schneider[2], James P. Best[1] and Gerhard Dehm[1,*]

[1] Max Planck Institute for Sustainable Materials, 40237 Düsseldorf, Germany

[2] Materials Chemistry, RWTH Aachen University, 52074 Aachen, Germany

*Corresponding authors: a.kanjilal@mpie.de, g.dehm@mpie.de

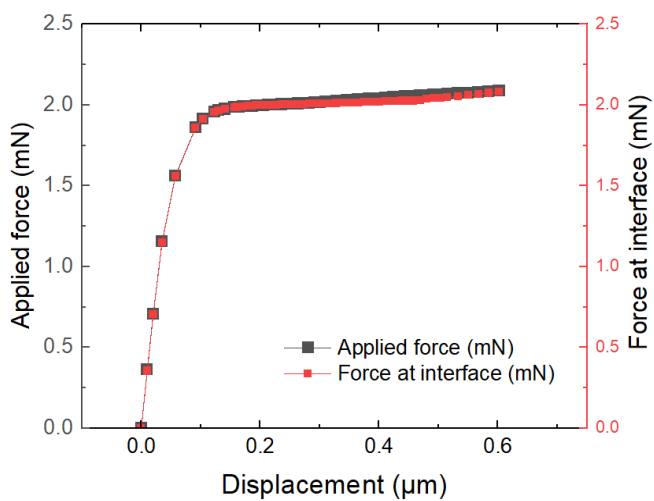

**Fig. S1**: Comparison of applied force on the $CaMg_2$ intermetallic with force at the interface obtained from finite element modeling based on which the interface shear stress is calculated.

## Fabrication of Ca-Mg films and identification of the region with $CaMg_2$ intermetallic

The bilayered model thin film system was deposited by combinatorial magnetron sputtering using direct current [1]. The deposition was carried out in a high vacuum chamber, pumped to a base pressure of $2 \times 10^{-5}$ Pa at a deposition temperature of 100 °C, with elemental targets of Mg (99.95%) and Ca (99.5%). The (001) Si substrate, located at a distance of 10 cm to the plasma sources, was kept electrically floating during deposition conducted in an Ar pressure of 0.5 Pa. Since the Ca oxidizes spontaneously upon atmosphere exposure, the Ca target was received in mineral oil. Prior to experimentation the target was cleaned and conditioned following procedures outlined in [1].

Onto the first pure Mg layer, sputtered at 200 W for 30 min, the Ca-Mg layer was grown. While the Mg target, oriented at an angle of 0° relative to the substrate normal, was maintained at 200 W, the Ca target, mounted at an inclination angle of 45° relative to the substrate normal, was co-sputtered at an applied power of 160 W for 30 min. To identify the region of interest corresponding to single-phased $CaMg_2$ with a ratio of Mg/Ca = 2 in the bilayered thin film used in this study, single-layered Ca-Mg combinatorial thin film was also deposited in a separate experiment with similar deposition parameters of applied powers of 200 and 160 W for Mg and Ca, respectively, for 15 min at a deposition temperature of 100 °C. Since Ca-Mg single-layered thin film was deposited with the same deposition parameters as that of the bilayered film, the chemical compositions and phases formed correspond to the same physical positions on both samples. To prevent impurity incorporation into the thin film model system by post deposition atmosphere exposure, a ~100 nm thick protective Mg capping layer was deposited on the graded Ca-Mg top layer at 200 W for 30 s, without intentional heating [1] (see **Fig. S2a** for schematic representation of the bilayered film).

Correlative chemical and structural analyses were performed on the single layered Ca-Mg film by energy-dispersive X-ray spectroscopy (EDX) and X-ray diffraction (XRD) to identify the region where $CaMg_2$ intermetallic formed. EDX measurement was performed using an EDAX Genesis 2000 analyzer in a JEOL JSM 6480 scanning electron microscope (SEM, JEOL Ltd., Tokyo, Japan). The acceleration voltage was set to 8 kV and the measurement time was 120 s at a magnification of 1000X. Furthermore, structural analysis of the film was performed using a Bruker D8 General Area Detection Diffraction System (GADDS, Bruker Corporation, Billerica, MA, USA) with Cu Kα radiation. The voltage and current settings were 40 kV and 40 mA, respectively. The angle of incidence was kept fixed at 15° whereas the 2θ range was 15° to 75°. Based on diffraction patterns and Mg/Ca ratios, calculated from the EDX analysis of the single-layered Ca-Mg thin film along the Mg gradient, the region of interest was chosen to correspond to single-phased $CaMg_2$ with a Mg/Ca = 2 as shown in **Fig. S2b**. The same coordinate was selected on the bilayered thin film, from which a cross-section region was extracted and its chemical composition was verified using EDX linescan as shown in **Fig. 2** of the main text.

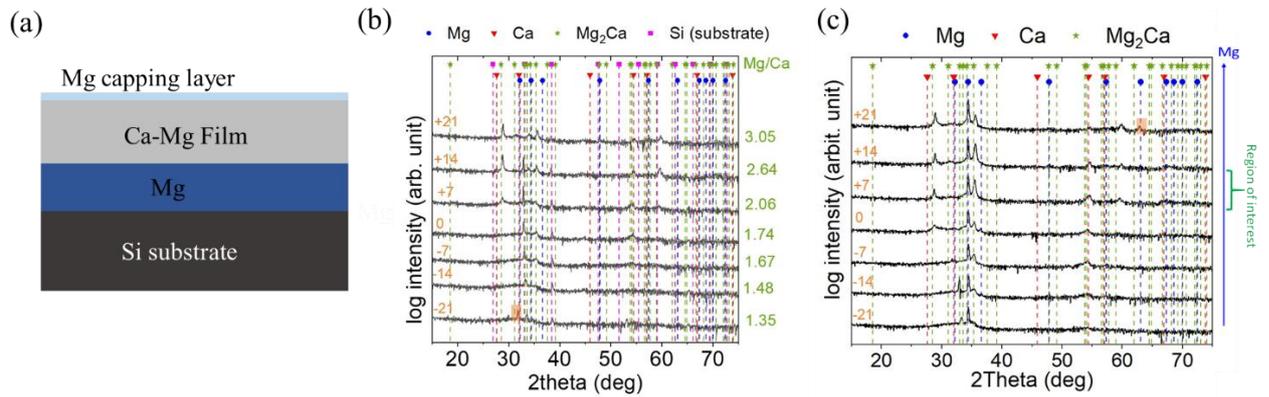

**Fig. S2**: (a) Schematic representation of Ca-Mg combinatorial thin film on Mg, (b) X-Ray diffraction patterns of the single-layered Ca-Mg thin film, with Mg/Ca ratio calculated based on EDX measurements, to identify the region of interest where $CaMg_2$ phase formed. Orange values are the distance to the middle of the sample in mm, (c) X-Ray diffraction patterns of the bilayered Ca-Mg film showing the stoichiometric $CaMg_2$ phase formed in the same region of interest as the single layered film.

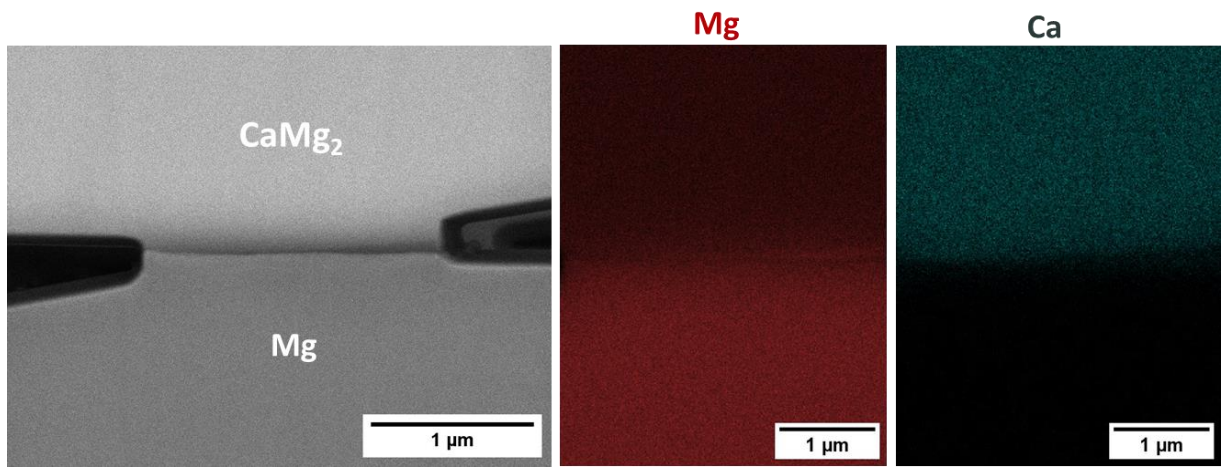

**Fig. S3**: SEM-EDS composition analysis across the sheared $CaMg_2$-Mg interface after testing shows the distribution of Mg and Ca across the interface of the sheared specimen indicating that shearing occurred near the interface.